\documentclass[prl,twocolumn]{revtex4}
\usepackage{amsfonts,amssymb,amsmath}            
\usepackage[]{graphics,graphicx,epsfig}            
\usepackage{amsthm}
\def\identity{\leavevmode\hbox{\small1\kern-3.8pt\normalsize1}}

\newcommand{\ket}[1]{\left | #1 \right\rangle}
\newcommand{\bra}[1]{\left \langle #1 \right |}
\newcommand{\half}{\mbox{$\textstyle \frac{1}{2}$}}

\newcommand{\Tr}{\text{Tr}}
\newcommand{\braket}[2]{\left\langle #1|#2\right\rangle}
\newcommand{\proj}[1]{\ket{#1}\bra{#1}}
\renewcommand{\epsilon}{\varepsilon}
\begin{document}

\title{The Capabilities of a Perturbed Toric Code as a Quantum Memory}
\date{\today}

\author{Alastair \surname{Kay}}
\affiliation{Centre for Quantum Technologies, National University of Singapore, 
			3 Science Drive 2, Singapore 117543}
\affiliation{Keble College, Parks Road, Oxford, OX1 3PG, UK}
\begin{abstract}
We analyze the effect of typical, unknown perturbations on the 2D toric code when acting as a quantum memory, incorporating the effects of error correction on read-out. By transforming the system into a 1D transverse Ising model undergoing an instantaneous quench, and making extensive use of Lieb-Robinson bounds, we prove that for a large class of perturbations, the survival time of stored information grows at least logarithmically with the system size. A uniform magnetic field saturates this scaling behavior. We show that randomizing the stabilizer strengths gives a polynomial survival time with a degree that depends on the strength of the perturbation.
\end{abstract}

\maketitle
The theories of quantum error correction and fault tolerance prove that quantum information can be stored despite the deleterious effects of external noise and experimental imperfection \cite{ft}. However, the massive overheads induce significant practical obstacles \cite{knill}. In contrast, classical memories are stable without active error correction. Can quantum information also be passively protected? To this end, quantum memories \cite{kitaev_fault-tolerant_2003,dennis-2002}, in which a qubit is encoded in the degenerate ground space of a Hamiltonian, have generated significant interest recently. One should design the Hamiltonian structure to prevent catastrophic accumulation of errors such that, after storage, a single round of error correction correctly returns the initial state. Crucially, the aim is to achieve storage times that scale with the system size.

The toric code of $2N^2$ qubits in 2D \cite{kitaev_fault-tolerant_2003,dennis-2002} is the prototypical proposal of such a system. While there is strong evidence that it is not a good memory at any finite temperature \cite{alicki-2008a,pacman1}, studies can yield significant insights for other types of noise, such as the imperfect implementation of an experiment at zero temperature, i.e.\ the effects of {\em unknown} static perturbations to the toric code Hamiltonian. The topological phase of the toric code is robust against perturbations \cite{kitaev_fault-tolerant_2003,hastings,specialcases}, meaning the degeneracy is only lifted to an exponentially small degree (in $N$), provided the perturbation strength $\delta\ll\Delta$, the unperturbed gap. States stored in this ground state space take exponentially long to dephase. If the perturbation is unknown, the challenge is to encode in this space, which may be achieved via an adiabatic path \cite{hamma,kay:prep}. At best, the final state is subject to a finite density of anyonic excitations which perturbations can easily propagate into logical errors \cite{kay:prl,nando}, although randomizing the weights in the unperturbed Hamiltonian induces Anderson localization and should reduce propagation in almost all cases \cite{pacman2}, while leaving the worst-case scaling unchanged \cite{nando}. 

Non-adiabatic methods, assisted by error correction, focus on accurately preparing the toric code. If this is not the ground state of the perturbed Hamiltonian, after some time the evolution may mask the stored information. In \cite{nando,kay:prl} pathological local perturbations ($\delta\ll \Delta$) showed the worst case survival time is no better than $O(\delta^{-1}\log(N))$. This paper considers a more typical experimental affliction, that of a uniform magnetic field. This is achieved by transforming the model into a parallel set of 1D transverse Ising chains subject to an instantaneous quench. While this model has previously been analyzed in the thermodynamic limit \cite{transverse}, we study the behavior for large, but finite, system sizes $N$, showing that the stored data is stable for times $\sim\Delta^2\delta^{-3}\log N$. 
We also analyze a system with randomized strengths, proving a polynomial survival time.

\begin{figure}
\begin{center}
\includegraphics[width=0.45\textwidth]{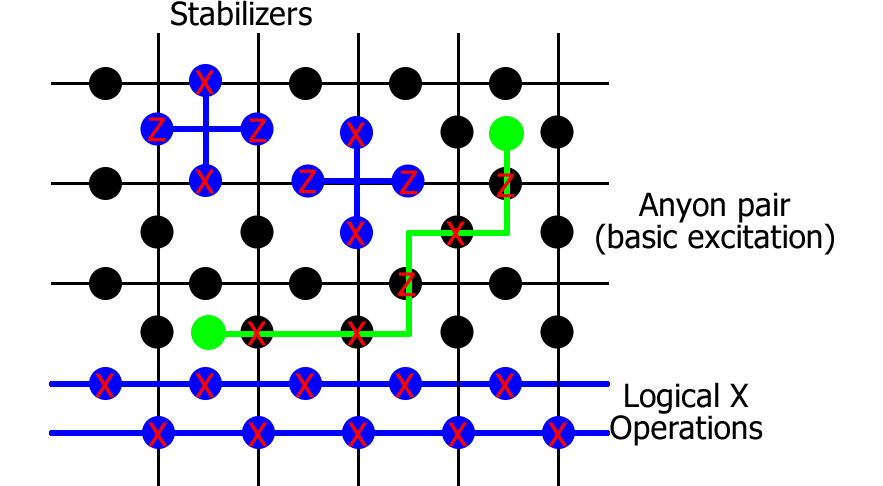}
\end{center}
\vspace{-0.5cm}
\caption{The 4-body stabilizers and logical $X$-rotations of the toric code. Anyons are propagated horizontally by $X$ rotations and vertically by $Z$ rotations.} \label{fig:toric}
\vspace{-0.5cm}
\end{figure}

{\em The Toric Code} is defined for an $N\times N$ periodic square lattice with a qubit placed in the middle of every edge, as depicted in Fig.\ \ref{fig:toric}. The Hamiltonian is a sum of 4-body commuting terms, $[K_n,K_m]=0$,
$$
H=-\sum_{n=1}^{2N^2}\Delta_nK_n.
$$
These terms $K_n$ are typically $ZZZZ$ on a face or $XXXX$ around a vertex, where $X$ and $Z$ are the standard Pauli matrices. There is a 4-fold degeneracy in the ground state space (defined by $K_n\ket{\psi}=\ket{\psi}$), allowing the encoding of two qubits. We consider a rotated version (apply Hadamards along every second row) in which every stabilizer term is the same, $XZZX$ \footnote{The code is identical to Wen's \cite{wen}, up to boundaries.}. In addition to making the model translationally invariant (if $\Delta_n=\Delta>0$), the logical $Z$ ($X$) rotations for the encoded qubits are two inequivalent columns (rows) of $Z$ ($X$) operators. The Hamiltonian is subsequently subject to a perturbation
$$
V=\sum_{n=1}^{2N^2}\delta_nX_n.
$$
Our aim is to determine times and field strengths such that, with high probability, error correction on the state $e^{-it(H+V)}\ket{\psi_0}$ produces a logically $X$ rotated state, $\ket{\psi_1}$ ($\hbar=1$). This is the time at which, for that model of error correction, the stored data is unreliable.

To locate errors, we measure the stabilizers, and perform minimum weight perfect matching (MWPM) to determine how to eliminate them. We assume that this process is perfect, otherwise the effectiveness, and hence storage time, is reduced. Since MWPM is difficult to analyze, we make use of a special property of the chosen perturbation. Specifically, a given $X_n$ commutes with all the stabilizers, except the two positioned to the left and right of the site $n$. Consequently, all the excitations of the original code due to the perturbation manifest themselves along horizontal lines. Every row must have an even number of errors. The chance of this happening by accident is vanishingly small, implying that the perturbation is just creating errors along the rows. Hence, we perform error correction by using MWPM independently along each row. When suitable, this method is strictly stronger than the full 2D error correction. 
If each row has a logical error with probability $p$, a logical error arises overall with a probability 
$
\half(1-(1-2p)^N).
$
For large $N$, any finite $p$ is destructive.

{\em The transverse Ising model:} The row-like feature that the perturbation induces means that the spectrum of the Hamiltonian, along each row, is equal to that of the transverse Ising model with periodic boundary conditions \cite{wen2}
$$
H_I=-\sum_{n=1}^N\Delta_nZ_nZ_{n+1}+\sum_{n=1}^N\delta_nX_n.
$$
A term $X_n$ commutes with all $ZZ$ terms other than $Z_{n-1}Z_n$ and $Z_nZ_{n+1}$.
The bit-flip distribution along a particular row of $e^{-it(H+V)}\ket{\psi_0}$ relative to $\ket{\psi_0}$ is the same as that on $e^{-itH_I}\ket{0}^{\otimes N}$. The equivalent error correction measures the stabilizers $Z_nZ_{n+1}$, and minimizes the number of flips to return the system to a $+1$ eigenstate of the stabilizers. Error correction fails if there is a finite probability that more than half the qubits have flipped, i.e.\ if $m$, the value of the magnetization
$$
M=\frac{1}{N}\sum_{n=1}^NZ_n,
$$
after projection onto the stabilizer space, is negative. We wish to argue that certain values of 
$$
\langle M\rangle=\tfrac{1}{N}\bra{0}^{\otimes N}e^{iH_It}\sum_nZ_ne^{-iH_It}\ket{0}^{\otimes N}
$$
correspond to finite $p$. Lieb-Robinson bounds \cite{LRHas} show that at time $t$ correlations between terms separated by a distance $|i-j|>\!\!2vt$ are exponentially small (hence negligible), i.e.\ for any operators $A$ and $B$ separated by distance $d_{AB}$, there exist positive constants $C,\eta$,
$$
\|[e^{-iHt}Ae^{iHt},B]\|\leq C\|A\| \|B\|e^{\eta(vt-d_{AB})},
$$
 where $v\sim\delta=\max\delta_n$ is the speed of sound for the system. In this regime, $\langle Z_iZ_j\rangle\approx \langle Z_i\rangle\langle Z_j\rangle$ \cite{frank}, so for times $t\ll N/\delta$, almost all the variables $Z_i$ are independent. Hence, Hoeffding's inequality applies to a good approximation \footnote{To avoid the approximation, one can apply Cantelli's inequality with little change in the ultimate results.} and, consequently, the probability that $m<0$ is finite iff $\langle M\rangle\sim O(1/\sqrt{N})$. This is the signature for failure of the memory.

{\em An upper bound:} Since $X_L=X^{\otimes N}$ is a conserved quantity of $H_I$, the initial state decomposes in terms of
$$
\ket{GHZ_{\pm}}=(\ket{0}^{\otimes N}\pm\ket{1}^{\otimes N})/\sqrt{2}.
$$
Given that $\{Z_n,X_L\}=0$ for all $n$,
$$
\langle M\rangle=\tfrac{1}{N}\sum_n\text{Re}\bra{GHZ_-}e^{iH_It}Z_ne^{-iH_It}\ket{GHZ_+}.
$$
Bounding this by the absolute value and using $\ket{GHZ_-}=Z_n\ket{GHZ_+}$ reveals that
\begin{eqnarray*}
\langle M\rangle&\leq&\tfrac{1}{N}\sum_n\left|\bra{GHZ_+}e^{iZ_nH_IZ_nt}e^{-iH_It}\ket{GHZ_+}\right|,
\end{eqnarray*}
which is restricted to the $+1$ eigenspace of $X_L$. In terms of the Majorana fermions
\begin{equation*}
c_{2n-1}=\left(\prod_{m=1}^{n-1}X_m\right)Z_n \qquad c_{2n}=i\left(\prod_{m=1}^{n-1}X_m\right)Y_n,
\end{equation*} 
$H_I$ is bilinear, with terms $\half h_{nm}c_nc_m$, where $h$ is a $2N\times 2N$ matrix ($\ket{2N+1}=\ket{1}$)
\begin{eqnarray*}
h&=&\sum_{n=1}^N\delta_n\left(\ket{2n-1}\bra{2n}-\ket{2n}\bra{2n-1}\right) \\
&&-\sum_{n=1}^N\Delta_n\left(\ket{2n}\bra{2n+1}-\ket{2n+1}\bra{2n}\right),
\end{eqnarray*}
given that evolution is only in the $+1$ eigenspace of $X_L$. We use $h^{(0)}$ to denote the instance of $h$ with $\delta_n=0$ and $\Delta_n=1$.
The GHZ state is a projection onto the simultaneous eigenspace of fermion pairs,
\begin{eqnarray*}
\proj{GHZ_+}\!&=&\!\!\lim_{\beta\rightarrow 0}\frac{1}{2^N\cosh^N\beta}e^{\half\beta\sum_{n,m}h_{n,m}^{(0)}c_nc_m},
\end{eqnarray*}
which is a fermionic Gaussian state with covariance matrix $h^{(0)}$ \cite{bravyi},
$
\proj{GHZ_+}=\omega(h^{(0)}).
$
A state $\rho_1=\omega(M)$ evolves under a Hamiltonian $H=\half\sum_{n,m}h_{nm}c_nc_m$ as
$$
e^{-iHt}\rho_1e^{iHt}=\omega(e^{-2ht}Me^{2ht}).
$$
Furthermore, for two states $\rho_1=\omega(M_1)$ and $\rho_2=\omega(M_2)$,
$$
\Tr(\rho_1\rho_2)=\tfrac{1}{2^N}\sqrt{\det(M_1)\det(M_2+M_1^{-1})}.
$$
Applying these rules leads to the conclusion that
\begin{eqnarray}
\langle M\rangle&\leq&\tfrac{1}{N}\sum_n\left|\det(\half B_n-\half \identity)\right|^{\frac{1}{4}},	\label{eqn:mainresult}\\
B_n&=&e^{2ht}e^{-2h_nt}h^{(0)}e^{2h_nt}e^{-2ht}h^{(0)},	\nonumber
\end{eqnarray}
where
$
h_n=h(\delta_n\mapsto-\delta_n)
$
results from $Z_nH_IZ_n=H_I-2\delta_n X_n$ being bilinear in fermions. 



{\em Analysis:} Eq.\ (\ref{eqn:mainresult}) gives an upper bound -- if it evaluates to $1/\sqrt{N}$, the magnetization cannot be larger, and the memory is unreliable. To see that the bound is tight, consider the evolution of $\ket{GHZ_\pm}$. Initially, these states remain close to the original states, with a low density of bit flips described by $Q$. We need to examine the phase 
$$
\theta=\text{Arg}\left(\frac{\bra{GHZ_+}Q^\dagger e^{-iH_It}\ket{GHZ_+}}{\bra{GHZ_-}Q^\dagger e^{-iH_It}\ket{GHZ_-}}\right).
$$
If $\theta=0$ for all populated $Q$, then the bound is exact, so we want to argue that $\theta$ remains small. Now consider the evolution of the initial state $\ket{0}^{\otimes N}$. In a given $Q$-sector (assumed to be of low density), the state becomes
\begin{equation}
Q(\ket{GHZ_+}+e^{-i\theta}\ket{GHZ_-})/\sqrt{2}. \label{eqn:qstate}
\end{equation}
As $\theta$ increases towards $\pi/2$, it becomes a GHZ state. However, Lieb-Robinson bounds show \cite{frank} that a 1D local Hamiltonian with speed of sound $v$ cannot create an $N$-qubit GHZ state in a time less than $O(N/v)$. Thus, our bound on $\langle M\rangle$ reveals not only when the memory is certainly not stable, but also when it is stable.

Returning to the calculation of $\det(\half(B_n-\identity))$, since $B_n$ is unitary and $B_n=B_n^*$, the eigenvalue pairs take the form $e^{\pm i\theta_i}$ (with the possible exception of $e^{i\theta}=\pm 1$). So,
\begin{eqnarray*}
\det(\half(B_n-\identity))&=&\tfrac{1}{4^N}\prod_i(e^{i\theta_i}-1)(e^{-i\theta_i}-1)	\\
&=&\tfrac{1}{2^N}\prod_i(1-\cos\theta_i).
\end{eqnarray*}
Provided $|x|\leq 1$,
$
\half(1-x)\leq e^{-(x+1)/2},
$
which reveals that
$
\langle M\rangle\leq \frac{1}{N}\sum_n\exp\left(\frac{-1}{16}(\Tr(B_n)+2N)\right)
$. We will take the trace of $B_n$ using a basis $\ket{\theta_m^{\pm}}=(\ket{2m-1}\pm i\ket{2m})/\sqrt{2}$, which is a diagonal basis of $h^{(0)}$. Lieb-Robinson bounds again allow us to describe the maximum distance that the $\ket{\theta_m^{\pm}}$ can be propagated by $h$ for any model $\{\Delta_n,\delta_n\}$. Indeed, provided $|m-n|\gg \delta t$, 
$$
e^{-2ih_nt}e^{2iht}\ket{\theta_m^{\pm}}\approx \ket{\theta_m^{\pm}}
$$
because the propagating states never `see' the different coupling strength between $h_n$ and $h$, so this means that only $O(\delta t)$ of the $\ket{\theta_m^{\pm}}$ may not result in $-1$ for the trace.

\begin{figure}
\begin{center}
\includegraphics[width=0.45\textwidth]{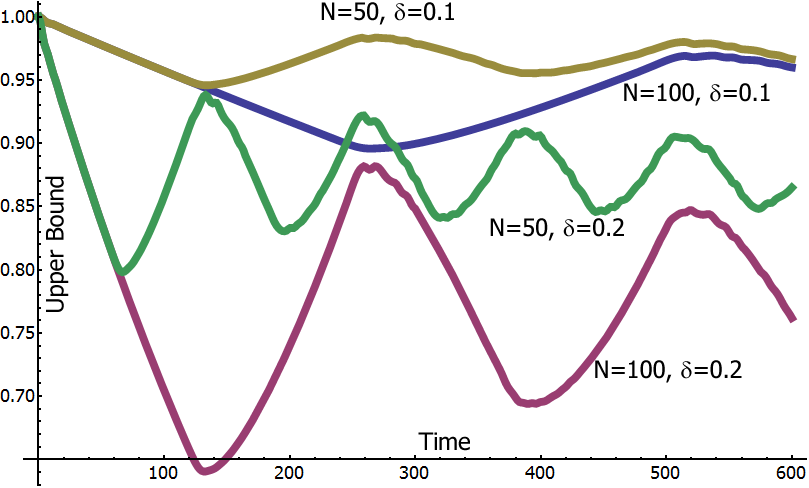}
\end{center}
\vspace{-0.5cm}
\caption{Time evolution of the upper bound to magnetization for several values of system size and perturbation strength -- $\Delta_n=1$ and $\delta_n$ is the same at every site.} \label{fig:bound}
\vspace{-0.5cm}
\end{figure} 

For those states $\ket{\theta_m^{\pm}}$ whose horizon contains the sites $2n-1$ and $2n$, $\bra{\theta_m^{+}}B_n\ket{\theta_m^{+}}$ simply measures the weight of the state $e^{-2ih_nt}e^{2iht}\ket{\theta_m^+}$ on the subspace $\ket{\theta_k^-}$. Moving to the interaction picture with respect to $h$, the only remaining evolution is $h_n-h$. Consider the eigenspace of $h$. If $\delta_n=0$, we just have the energies $\{\pm\Delta_n\}$, and the eigenstates are categorized into two bands by whether they are of positive or negative energy (corresponding to having support on one of the two subspaces $\ket{\theta_k^{\pm}}$). By adding in the $\{\delta_n\}$, there is additional coupling, both inter- and intra-band. Within the band, we completely solve for the new eigenvectors, treating inter-band coupling as a perturbation ($\Delta_{\text{min}}\gg\delta$). The positive band has energies in the range $[\Delta_{\text{min}}-\delta,\Delta_{\text{max}}+\delta]$. The first order correction to the eigenvectors in a perturbative expansion is $O(\delta/\Delta_{\text{min}})$. Whilst $h_n-h$ couples between all of these eigenvectors, for times $t\gg\Delta_{\text{min}}^{-1}$ the rotating wave approximation averages away the coupling between the two bands. Hence, if a state has support on one particular band, it remains in that band. We have just argued however, that these bands correspond to being in the two different spaces $\ket{\theta_k^+}$ and $\ket{\theta_k^-}$, except for $O(\delta/\Delta_{\text{min}})$ corrections to the amplitude. Hence no more than $O((\delta/\Delta_{\text{min}})^2)$ weight is lost to the other subspace during the evolution. Thus,
$$
\Tr(B_n)\leq -(2N-\beta\delta t)-(\beta-\alpha(\delta/\Delta_{\text{min}})^2)\delta t
$$
where $\alpha,\beta$ represent undetermined constants from our perturbation argument and Lieb-Robinson bound. If $\alpha\delta^3 t/\Delta_{\text{min}}^2\ll 1$, then we conclude that no model will ever have a value of $\langle M\rangle$ that scales worse than
$$
e^{-\alpha\delta^3 t/(16\Delta_{\text{min}}^2)}\approx 1-\alpha\delta^3 t/(16\Delta_{\text{min}}^2).
$$
For sufficiently weak perturbations, the magnetization drops linearly with time. For stronger perturbations and sufficiently long times (which always exist if $\alpha(\delta/\Delta_{\text{min}})^2 N>16$), the magnetization decays exponentially towards zero, and the information is unreliable once it drops below $1/\sqrt{N}$. Hence, the system presents a survival time of $\Omega(\delta^{-3}\Delta_{\text{min}}^2\log N)$. Every model $\{\Delta_n,\delta_n\}$ characterized by $\delta/\Delta_{\min}$ certainly survives this long. Not all models propagate information in a way that saturates Lieb-Robinson bounds. However, the uniform field case $\Delta_n=1, \delta_n=\delta$ has a finite group velocity and is hence expected to exhibit this scaling. Numerical confirmation can be seen in Fig.\ \ref{fig:bound}. Although this is a relatively weak effect at small system sizes ($\delta/\Delta_{\min}=0.2$ has a critical system size of about $N=600$, at which point survival is for a time of approximately $1500\Delta_{\min}^{-1}$), the scaling is so weak that it could easily dominate other timescales as one approaches mesoscopic systems.

\begin{figure}
\begin{center}
\includegraphics[width=0.45\textwidth]{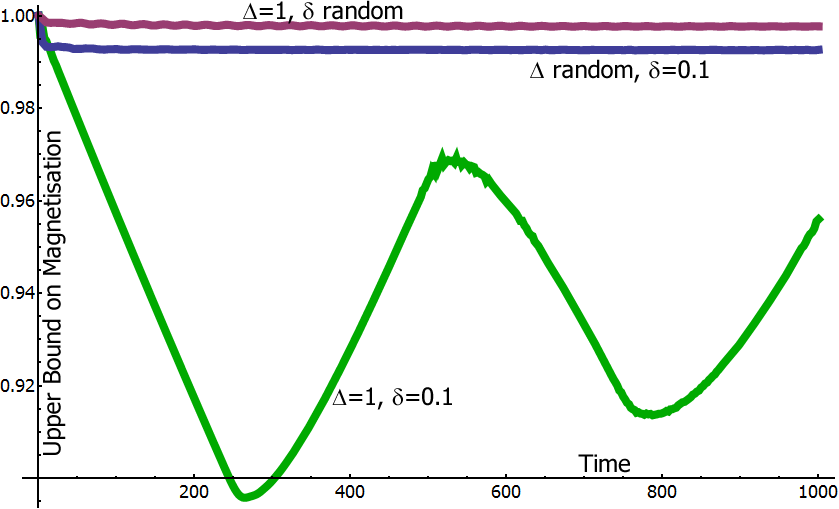}
\end{center}
\vspace{-0.5cm}
\caption{Comparison of 3 cases ($N=100$): $\Delta=1, \delta=0.1$; $\delta=0.1$, $\Delta_n\in[0.5,1]$; $\Delta=1$, $\delta_n\in[-0.1,0.1]$. Random cases are sampled uniformly and averaged over 100 instances.} 
\vspace{-0.5cm}
\label{fig:random}
\end{figure}

{\em Random Systems:} In random systems, the propagation of information is suppressed compared to that predicted by Lieb-Robinson bounds, enabling much longer storage times. To this end, we aim to induce Anderson localization by randomizing either the $\Delta_n$ (by designing the system) or the $\delta_n$ (as might arise naturally), concentrating on the former and maintaining the assumptions that $\Delta_n>0$ and $\Delta_{\min}\gg\delta$. Fig.\ \ref{fig:random} already confirms such expectations numerically. Localization embodies two concepts: {\em spectral localization} (SL) requires that every eigenvector $\ket{\lambda_k}$ of $h$ has exponentially decaying tails, i.e.\ there exist positive constants $C_k, \eta_k$ independent of $N$ such that $|\braket{n}{\lambda_k}|\leq C_ke^{-\eta_k|n-k|}$ \footnote{For clarity of exposition, we have suppressed the periodic boundary conditions, and how SL should be restated for the two-band structure of our model \cite{new_loc}.}, whereas {\em dynamical localization} (DL) requires that $|\bra{m}e^{-iht}\ket{n}|$ is exponentially bounded in $|m-n|$. While DL, which implies SL, holds for almost all instances of the $XX$ model with random magnetic field \cite{new_loc2}, to our knowledge, neither has been proven for the transverse Ising model. However, the interaction within each band of our model is identical to the $XX$ model. Hence, one can readily show that the perturbative inter-band coupling leaves SL intact. \cite{new_loc} showed that if DL can be proven for the matrix $h$, a Lieb-Robinson bound with finite horizon follows for $H$. The same argument applied to SL shows that there exist positive coefficients $c,C'$ and $\eta'$ such that
$$
\|[e^{-iHt}Ae^{iHt},B]\|\leq \min(c|t|,C')N\|A\| \|B\|e^{-\eta'd_{AB}}.
$$
Hence, the system exhibits a logarithmic light cone akin to that of \cite{tobias}. Replacing all three applications of Lieb-Robinson bounds in our previous analysis, we conclude that sufficiently large systems are stable for a time at least $t\sim c^{-1}N^{\Omega(\Delta_{\min}^2\delta^{-2})}$ for almost all instances. 

{\em Conclusions:} When subject to an unknown perturbation along a particular field direction ($X$ or $Z$), then, in every instance, information stored in the degenerate ground space of the toric code is stable for a time $\Omega(\delta^{-3}\Delta_{\text{min}}^2\log N)$, comparable to the worst-case scaling for any perturbation constructed in \cite{nando}. The analysis has been confirmed numerically for system sizes of up to $2\times 10^6$ qubits. We have also examined randomized stabilizer strengths \cite{pacman2}, improving upon previous analyses by accurately encapsulating the effects of particle creation in the perturbation, and of error correction. This has enabled us to show that the survival time almost always scales at least polynomially in $N$ for finite system sizes, which is a consideration entirely absent from \cite{pacman2}, and yet is of central importance. In the future, proving dynamical localization of the 1D transverse Ising model could show that storage is stable for arbitrary times.


While we have argued that the use of 1D error correction is well justified, how would this compare to the full 2D error correction? If the error density remains low, typical occurrences will be single localized errors. Where the 1D error correction has a threshold probability of $50\%$ for these errors, we have numerically estimated a threshold for the 2D error correction to be $22\%$ (as compared to $11\%$ \cite{dennis-2002} for arbitrary local errors), removing all $N$ dependence from our proof, and reducing our lower bounds on the survival time to a constant.


{\em Acknowledgments:} This work is supported by the National Research Foundation \& Ministry of Education, Singapore. The author thanks Albert Werner for useful discussions.

\end{document}